\documentclass[aps,prd,preprintnumbers,floats,epsfig,nofootinbib,twocolumn,nofootinbib]{revtex4-1}

\usepackage{slashed}
\usepackage{amsmath,amssymb,amsfonts,mathtools}
\usepackage{graphicx,color}
\usepackage{epsfig}
\usepackage{subfigure}
\usepackage{epsfig}
\usepackage{dcolumn}
\usepackage{bm}
\usepackage{color}
\usepackage{ulem}
\usepackage{enumitem}
\usepackage[colorlinks,
            linkcolor=black,
            anchorcolor=black,
            citecolor=black
            ]{hyperref}

\begin{document}

\baselineskip=15pt


\title{Flavor-changing Majoron interactions with leptons}

\author{Yu Cheng$^1$\footnote{chengyu@sjtu.edu.cn}}
\author{Cheng-Wei Chiang$^{2,3}$\footnote{chengwei@phys.ntu.edu.tw}}
\author{Xiao-Gang He$^{2,3}$\footnote{hexg@phys.ntu.edu.tw}}
\author{Jin Sun$^1$\footnote{019072910096@sjtu.edu.cn}}

\affiliation{${}^{1}$Tsung-Dao Lee Institute, and School of Physics and Astronomy, Shanghai Jiao Tong University, Shanghai 200240, China}
\affiliation{${}^{2}$Department of Physics, National Taiwan University, Taipei 10617, Taiwan}
\affiliation{${}^{3}$Physics Division, National Center for Theoretical Sciences, Taipei 10617, Taiwan}

\begin{abstract}
When the Standard Model Higgs sector is extended with a complex singlet that breaks global lepton number symmetry spontaneously, a massless Goldstone boson called the Majoron $J$ arises.  In addition to increasing Higgs invisible decay through mixing, the Majoron can generally have flavor-changing interactions with fermions.
We find that type-III seesaw model poses such interesting properties with both charged leptons and neutrinos.
This opens up new channels to search for the Majoron.  We use the experimental data such as muonium-anti-muonium oscillation and flavor-changing neutrino and charged lepton decays to put constraints on the couplings.  As a novel way to reveal the chiral properties of these interactions, we propose an experimentally measurable polarization asymmetry of flavor-changing $\ell \to \ell' J$ decays.
\end{abstract}

\maketitle

\section{Introduction}
\label{sec:Introduction}

Since the discovery of the 125-GeV Higgs boson, it is an intriguing question whether there exists other elementary scalar bosons in nature.  A class of models with an economical extension of the Higgs sector in the Standard Model (SM) involve the introduction of a $SU(2)_L \times U(1)_Y$ complex singlet $S$ that induces the breakdown of global lepton number symmetry $U(1)_L$.  Such models are of primary interest because they can generate neutrino mass~\cite{Chikashige:1980qk} and house a Goldstone boson, generically called the Majoron $J$, from the spontaneous symmetry breaking triggered by the vacuum expectation value (VEV) of $S$.  Due to the singlet nature of $S$, the Majoron has no or very weak couplings with most SM particles, and can readily evade stringent constraints on massless boson searches.  The Majoron also has a lot of implications in astrophysics and cosmology~\cite{Kim:1986ax, Calibbi:2020jvd, Berezhiani:1990wn, Hirsch:2009ee}. While most existing phenomenological studies rest on its flavor-conserving interactions with fermions, we focus in this work on its flavor-changing couplings as constrained by laboratory experiments, and propose a new experimentally measurable polarization asymmetry to study the chiral structure of the couplings.

In general, the Majoron can induce flavor-changing interactions even in the simplest Type-I seesaw model if the heavy neutrino mass is generated by the VEV of $S$~\cite{Chikashige:1980qk}.  In this scenario, the Majoron only interacts with the neutrinos and the Higgs boson at tree level.  In particular, the Majoron couples to the Higgs boson through mixing, inducing the Higgs decay to two Majorons manifested as invisible decay.
There are also simple models where the Majoron can interact with other fermions.
In this work, we use Type-III seesaw model~\cite{ss-III}, with the VEV of a singlet to provide the heavy seesaw mass, as a simple explicit example to demonstrate the possibility of having flavor-changing interactions between the Majoron and the charged leptons. Such interactions alone will induce, for example, the $\mu \to e J$ decay, as considered a long time ago~\cite{sokolov}.  Such interactions open up new channels to search for the effects of Majoron, and may have novel implications on flavor-changing lepton decays.  In particular, if one can measure the helicity of charged lepton in the final state, the proposed polarization asymmetry for the flavor-changing $\ell \to \ell' J$ decays can reveal the chiral properties of the interactions without the need to know the initial-state lepton polarization.  With the same Majoron couplings  to neutrinos as in Type-I seesaw model, Type-III seesaw model serves as a more general framework, which we will examine in this paper.

The structure of this paper is organized as follows.  In Section~\ref{sec:Majoron in Type-III seesaw model}, we review the Type-III seesaw model and use it as an example to motivate the Majoron with flavor-changing couplings with fermions.  Section~\ref{sec:Constraints} shows major experimental constraints on the Majoron and thus the bounds on the ratio of couplings and the Majoron decay constant.  In Section~\ref{sec:Polarization Asymmetry}, we propose a new observable, the polarization asymmetry, in the $\ell_i \to \ell_j J$ decays to probe the chiral nature of the Majoron interactions with the fermions, and study the experimental feasibility.  Section~\ref{sec:Conclusion} summarizes our findings.

\section{Majoron in Type-III seesaw model}
\label{sec:Majoron in Type-III seesaw model}

In Type-III seesaw model, besides the three generations of left-handed lepton doublets $L_{iL}: (1,2,-1/2,1)$ and the right-handed charged leptons $E_{iR }: (1,1,-1,1)$ in the SM, there are also three generations of right-handed lepton triplets $\Sigma_{iR}: (1,3,0,1)$, where the numbers in the parentheses indicate their $SU(3)_C$, $SU(2)_L$, $U(1)_Y$ and $U(1)_L$ quantum numbers, respectively. The component fields of $\Sigma_R$ and its charge conjugated fields $\Sigma_L$ are as follows:
\begin{eqnarray}
\Sigma_R
= \left (\begin{array} {cc}
{\Sigma_L^{0\; c}\over \sqrt{2}}&\;\Sigma^{-\; c}_L\\
\Sigma^{+ \; c}_L&\;-{\Sigma^{0\;c}_L\over \sqrt{2}}
\end{array}
\right )
,~~
\Sigma_L
= \left ( \begin{array} {cc}
{\Sigma_L^0\over \sqrt{2}}&\;\Sigma^+_L\\
\Sigma^-_L&\;-{\Sigma^0_L\over \sqrt{2}}
\end{array}
\right ).
\end{eqnarray}
We will rename them as
$\nu_R = \Sigma_L^{0\; c}$, $\psi_L = \Sigma^-_L$ and $\psi_R = \Sigma^{+\;c}_L$.
The Higgs sector contains the usual Higgs doublet $H = (h^+, (v+ h + i I)/\sqrt{2})^T: (1,2,1/2, 0)$ with $h^+$ and $I$ to be ``eaten'' by the $W^+$ and $Z$ bosons, and an additional Higgs singlet $S = (1/\sqrt{2})(v_s + h_s + i I_s): (1,1,0,-2)$.  The Yukawa interactions involving the leptons conserve the global lepton number and are given by
\begin{align}
\begin{split}
{\cal L}_Y
=&
- \overline{ \ell_\ell} Y_e H E_R
- \overline{ \ell_\ell} \sqrt{2} Y_\nu \Sigma_R \tilde H - {1\over 2}
{\rm Tr}\,\overline{ \Sigma_R^{c}} Y_s S \Sigma_R
\\
&\qquad
+ \mbox{H.c.}
~,
\end{split}
\end{align}
where $\tilde H = i\sigma_2 H$ with $\sigma_i$ being the Pauli matrices.
Since the non-zero $v_s$ breaks the global lepton number, the imaginary component of $S$ emerges as a Goldstone boson widely known as the Majoron $J$, with $J= I_s$.

The Lagrangian terms relevant to charged lepton and neutrino masses and the interactions of $J$ to fermions are given by
\begin{eqnarray}
&&
- {1\over 2} \left( \overline{\nu_L}, \overline{\nu^{ c}_R} \right) M_\nu
\begin{pmatrix}
\nu^c_L\\\nu_R
\end{pmatrix}
 -(\overline{ E_L}, \overline{ \psi_L}) M_c \left ( \begin{array}{c}
E_R\\
\psi_R
\end{array}
\right )\nonumber\\
&&
- i  {J\over 2 f_J } \left [ \overline{\nu^c_R} M_R \nu_R
 -2  \overline{ \psi_L} M_R \psi_R \right ]+ \mbox{H.c.}
\;,\;\;
\end{eqnarray}
with
\begin{eqnarray}
&&M_\nu = \left (\begin{array}{ll}
0&\;\;M_D\\
M^T_D&\;\;M_R
\end{array} \right )
~,~~
M_c = \left ( \begin{array}{cc}
M_e&\;\; \sqrt{2} M_D\\
0&\;\;M_R
\end{array}
\right )\;,\nonumber\\
&&M_e =  {Y_e v\over \sqrt{2}}
~,~~
M_D =  {Y_\nu v\over \sqrt{2}}
~,~~
M_R =  {Y_s v_s\over \sqrt{2}}\;,
\end{eqnarray}
where $f_J = v_s$ is the Majoron decay constant that sets the seesaw scale.
The Majoron interaction terms in the form of derivative couplings are
\begin{eqnarray}
\frac{\partial_\mu J}{2 f_J} \left [\overline{\nu_L} \gamma^\mu \nu_L +  \overline{\nu_R} \gamma^\mu \nu_R - 2 \overline{ \psi_L} \gamma^\mu \psi_L\right ]
~.
\end{eqnarray}

The mass matrix $M_\nu$ and $M_c$ can be diagonalized in the forms $\hat M_\nu = V^\nu M_\nu V^{\nu T}$ and $M_c = {V^{e\,L}}^\dagger \hat M_c V^{e\,R}$. Here $V^\nu$, $V^{e \, L(R)}$ are $6\times 6$ unitary matrices.
Breaking $V^\nu$ into blocks of $3\times 3$ matrices, we have
\begin{eqnarray}
V^\nu = \left( \begin{array}{ll}
V^\nu_{LL}\;\;&V^\nu_{LR}\\
\\
V^\nu_{RL}\;\;&V^\nu_{RR}
\end{array}
\right)
,~
V^{e\;L(R)} = \left( \begin{array}{ll}
V^{e\;L(R)}_{LL}\;\;&V^{e\;L(R)}_{LR}\\
\\
V^{e\;L(R)}_{RL}\;\;&V^{e\;L(R)}_{RR}
\end{array}
\right)
.
\end{eqnarray}
Note that $\hat M_\nu$ is diagonalized in the basis $(\nu_L, \nu^c_R)$, therefore $ \overline{ \nu_L} \gamma^\mu \nu_L +  \overline{ \nu_R} \gamma^\mu \nu_R = \overline{ \nu_L} \gamma^\mu \nu_L -  \overline{ \nu^c_R} \gamma^\mu \nu^c_R$ after rotation will not be diagonal and lead to flavor-changing $J$ interactions with neutrinos. Also, as $V_{LL,LR, RL, RR}$ are not separately unitary, there are in general flavor-changing interactions induced in the charged lepton sector.  One can reduce to Type-I seesaw model by dropping the Majoron interactions with charged leptons.

Working in the basis where $M_e$ and $M_R$ are diagonal, one can approximate~\cite{Abada:2007ux,He:2009tf} $V_{LL} = (1-\epsilon/2)V_{PMNS}$ with $\epsilon = Y_DM_R^{-2} Y^\dagger_D v^2/2$. A global fit finds that the matrix elements in $\epsilon$ are ${\cal O}(10^{-3})$~\cite{Fernandez-Martinez:2016lgt}. Therefore, the couplings $V^\nu_{LL}{V^\nu_{LL}}^\dagger$ are allowed at the level of $10^{-3}$.  If different singlets are introduced for corresponding right-handed neutrinos to have different lepton numbers, one would change the Majoron couplings with light neutrinos to $V^\nu_{LL}X^\nu_R{V^\nu_{LL}}^\dagger$, with $X^\nu_R$ being a diagonal matrix of generally different entries~\cite{Sun:2021jpw}.  Individual off-diagonal couplings can now be much larger than $10^{-3}$ and should therefore be constrained by data. There are also constraints from mixing between heavy and light neutrinos, which can be independent of light neutrino mixing~\cite{He:2009ua}.

It is worth emphasizing that the Majoron generally also has flavor-changing interactions with charged leptons.  The sizes of the couplings are model-dependent and are {\it a priori} unknown.  We will treat them as theory parameters and constrain them using experimental data.  For this purpose, we generically write the Majoron couplings to the light charged leptons and neutrinos as
\begin{eqnarray}
{\partial_\mu J \over 2 f_J} \left [ \overline{ \ell_j} \gamma^\mu (c^{e\;ji}_V + c^{e\;ji}_A\gamma_5) \ell_i
+ \overline{ \nu_{Lj}} \gamma^\mu c^{\nu\;ji}_{L} \nu_{Li} \right ]
~,
\label{eq:derivative}
\end{eqnarray}
where $i$ and $j$ are flavor indices for the initial and final states, respectively. In Type-III seesaw model, $c^{e}_V=- c^{e}_A=-V^{e\;L}_{LR}{V^{e\;L}_{LR} }^\dagger$ and $c^{\nu}_L = V^{\nu}_{LL}{V^\nu_{LL}}^\dagger - {V^\nu_{LR}V^\nu_{LR}}^\dagger$.
For on-shell fermions, we get
\begin{eqnarray}
{i\over 2} J \left [\overline{\ell_j}  (g_{e1}^{ji} +  g_{e2}^{ji} \gamma_5 ) \ell_i
+ \overline{\nu_j} (g_{\nu1}^{ji} + g_{\nu2}^{ji} \gamma_5) \nu_i \right ]
~,
\label{eq:on-shell}
\end{eqnarray}
where $\nu$ denote the light neutrinos, and $g_{e1/e2}^{ji} = -(m_j c^{e\;ji}_{V/A} \mp  c^{e\;ji}_{V/A} m_i)/ f_J $, and $g_{\nu1/\nu 2}^{ji} =  ( c^{\nu\;ji}_{L} m_{\nu_i} \mp  m_{\nu_j} c^{\nu\;ji}_{L} )/2f_J$ with $m,\;m_{\nu}$ being the eigen-mass matrices of the charged leptons and light neutrinos, respectively.

The scalar potential in this model is given by
\begin{align}
\begin{split}
V(H, S)
=& -\mu^2 H^\dagger H + \lambda (H^\dagger H)^2 -\mu^2_s S^\dagger S
\\
& \qquad
+ \lambda_s (S^\dagger S)^2 + \lambda_{hs} (H^\dagger H)(S^\dagger S)
~.
\end{split}
\end{align}
Therefore, the Higgs boson naturally mixes with the real part of $S$ and couples with the Majoron through
\begin{eqnarray}
{1\over 2} \lambda_{hs} v  (h_1 \cos\theta + h_2 \sin\theta) J^2\;,
\end{eqnarray}
with $\tan 2\theta = \lambda_{hs} v v_s/(\lambda_s v^2_s - \lambda v^2)$, leading to the Higgs decay to Majorons which increases Higgs invisible width~\cite{Gelmini:1980re}.
Here we assume $h_1 \equiv h$ is the observed 125-GeV Higgs boson.

\section{Constraints}
\label{sec:Constraints}

Because of the flavor-changing Majoron interactions with neutrinos, decays of the type $\nu_i \to \nu_j J$ can occur, making neutrinos unstable. The decay width is given by $\Gamma(\nu_i \to \nu_j J) = (\Delta m^2_{ij})^3|c^{\nu\; ji}_L |^2/ (32 \pi m_i^3 f_J^2)$.  At present, without information of individual neutrino masses, the measured mass differences imply two mass orderings: normal hierarchy ($m_1 < m_2 < m_3$) and inverted hierarchy ($m_3 < m_1 < m_2$).  In principle, data on the lifetime-mass ratio, $\tau_i/m_i$, for neutrinos can constrain the parameters, yet current data~\cite{Zyla:2020zbs} do not give useful constraints.
For example, taking $m_1=0$ and $m_3=0$ respectively for the normal and inverted hierarchy cases, we have  $\tau_3/m_3 = 1.57\times 10^{43} (f^2_J/\mbox{GeV}^4)/\left( 0.91 |c^{\nu\; 23}_L |^2+ |c^{\nu\; 13}_L |^2 \right)$ and $\tau_2/m_2 = 1.77 \times 10^{46} (f^2_J/\mbox{GeV}^4)/|c^{\nu\; 12}_L |^2$ in the former case, and $\tau_2/m_2 = 1.55 \times 10^{43}(f^2_J/\mbox{GeV}^4)/\left( 2.59\times 10^{-5}\ |c^{\nu\; 12}_L |^2+ |c^{\nu\; 32}_L |^2 \right)$ and $\tau_1/m_1 = 1.65 \times 10^{43}(f^2_J/\mbox{GeV}^4)/|c^{\nu\; 31}_L |^2$ in the latter case.  These numbers are orders of magnitude above the current data~\cite{Zyla:2020zbs} if one demands $f_J$ to be as low as the weak scale and $c^{\nu\;ji}_{V,A}$ not to exceed order $O(1)$. For nonzero $m_1$ and $m_3$ cases, the situation gets worse.  It is thus clear that currently no constraints can be placed on neutrino-Majoron flavor-changing interactions.

As the Majoron can also have flavor-changing interactions with the charged leptons, much more severe constraints can be obtained from related processes.  Concentrating on charged lepton interactions, we will drop the superscript $e$ in $c^{e~ji}_{V,A}$ in the following discussions for notation simplicity.  We will consider three classes of constraints: (a) muonium-anti-muonium ($M$-$\overline M$) oscillation, (b) $\ell_i\to \ell_j J$ decays, and (c) $\ell_i \to \ell_j \ell_k \bar \ell_\ell$ decays.

\begin{table*}
\caption{Bounds on flavor-changing Majoron couplings with charged leptons.  Each bound is obtained by keeping only one type of interaction at a time.
\label{tab:constraints}}
\begin{ruledtabular}
\begin{tabular}{ccll}
&Process& Experimental input& Bound (in units of $f_J/\mbox{TeV}$) \\
\hline
I
& $M\to \overline{M}$
&${\rm P}<8.3\times 10^{-11}/S_B(B_0)$~\cite{Willmann:1998gd}
&$$\\
&$$
&$S_B(B_0)_{SS}=0.50$
&$|c_{V}^{\mu e }|<0.407$\\
&$$
&$S_B(B_0)_{PP}=0.9$
&$|c_{A}^{ \mu e}|<0.351$\\
&$$
&$S_B(B_0)_{(S \pm P) (S \pm P)}=0.35$
&$|c_{V/A}^{\mu e }|<0.444$\\
\hline
II
&$\mu\to e J$
&${\rm Br}<2.6\times 10^{-6}~  (90\%~{\rm CL})$~\cite{Jodidio:1986mz}
&$|c_{V/A}^{e \mu }|<3.64 \times 10^{-7}$ \\
&$\tau\to \mu J$
&${\rm Br}<5.7\times 10^{-3}~  (95\%~{\rm CL})$~\cite{Albrecht:1990zj}
&$|c_{V/A}^{\mu \tau}|<6.87 \times 10^{-4}$ \\
&$\tau\to e J$
&${\rm Br}<3.2\times 10^{-3}~  (95\%~{\rm CL})$~\cite{Albrecht:1990zj}
&$|c_{V/A}^{e \tau}|<5.11 \times 10^{-4}$ \\
\hline
III
&$\tau\to \mu e \bar \mu$
&${\rm Br}<2.7\times 10^{-8}~  (90\%~{\rm CL})$~\cite{Hayasaka:2010np}
&$\sqrt{|c^{\mu \tau}_{V/A}||c^{e \mu}_{V/A}|} < 0.379-0.405$ \\
&$\tau\to \mu e \bar e$
&${\rm Br}<1.8\times 10^{-8}~  (90\%~{\rm CL})$~\cite{Hayasaka:2010np}
&$\sqrt{|c^{e \tau}_{V/A}||c^{\mu e }_{V/A}|} < 0.353-0.355$ \\
&$\tau\to \mu \mu \bar e$
&${\rm Br}<1.7\times 10^{-8}~  (90\%~{\rm CL})$~\cite{Hayasaka:2010np}
&$\sqrt{|c^{\mu \tau}_{V/A}||c^{\mu e }_{V/A}|} < 0.346-0.349$ \\
&$\tau\to e e \bar \mu$
&${\rm Br}<1.5\times 10^{-8}~  (90\%~{\rm CL})$~\cite{Hayasaka:2010np}
&$\sqrt{|c^{e \tau}_{V/A}||c^{e \mu}_{V/A}|} < 0.346-0.347$\\
\hline
IV
&$(g-2)_e$
&$-(0.88 \pm 0.36) \times 10^{-12}$~ 
~\cite{Keshavarzi:2019abf}
&$|C_{A}^{e \mu}|<3.21,\;|C_{A}^{e \tau}|<0.782$\\
&$(g-2)_\mu$
&$(28.02 \pm 7.37) \times 10^{-10}$~ 
~\cite{Keshavarzi:2019abf}
&$|C_{V}^{\mu \tau}|<3.07$\\
\hline
V
&$\mu \to e \gamma$
&${\rm Br}<4.2\times 10^{-13}~  (90\%~{\rm CL})$~\cite{TheMEG:2016wtm}
&$\sqrt{|C_{V/A}^{e \tau}||C^{\tau \mu }_{V/A}|}<0.011$\\
&$\tau \to \mu \gamma$
&${\rm Br}<4.4\times 10^{-8}~  (90\%~{\rm CL})$~\cite{Aubert:2009ag}
&$\sqrt{|C_{V/A}^{\mu e }||C^{e \tau}_{V/A}|}<5.14 $\\
&$\tau \to e \gamma$
&${\rm Br}<3.3\times 10^{-8}~  (90\%~{\rm CL})$~\cite{Aubert:2009ag}
&$\sqrt{|C_{V/A}^{e \mu}||C^{\mu \tau}_{V/A}|}<4.78$\\
\end{tabular}
\end{ruledtabular}
\end{table*}

Case (a) is induced first by exchanging $J$ to produce the $(\bar \mu(c_V + c_A \gamma^5) e)^2$ operator which causes  muonium and anti-muonium to oscillate.  Including both $s$- and $u$-channel contributions and averaging over the spin-0 and -1 contributions, we have the oscillation probability
\begin{align}
{\rm P}(M\to \overline M)
= \frac{2\tau_\mu^2 \alpha^6 m_e^6}{\pi^2 f^4_J} \left ( |c_V^{\mu e}|^4+|c_A^{\mu e}|^4-|c_V^{\mu e}|^2|c_A^{\mu e}|^2\right )
.
\label{P}
\end{align}
Since no such oscillation is observed, experiments put a stringent bound on the  spin-0 and spin-1 muonium averaged oscillation probability, ${\rm P}(M\to \overline M)^{\rm exp} < 8.3\times 10^{-11}/S_B(B_0)$ in $B_0 = 0.1$~T~\cite{Conlin:2020veq,Willmann:1998gd}, where $S_B(B_0)$ is the magnetic field correction factor depending on the interaction type, used to describe the suppression of conversion in the external magnetic field $B_0$ due to the removal of degeneracy between corresponding levels in $M$ and $\bar{M}$.  The constraints based upon different correction factors for different interaction types are given in block I of Table~\ref{tab:constraints}.

The calculations for case (b) $\ell_i \to \ell_j J$ and case (c) $\ell_i \to \ell_j \ell_k \bar \ell_\ell$ are straightforward.  For case (b),  we will use the strongest experimental bounds available~\cite{Zyla:2020zbs} to constrain the parameters $|c_{V/A}^{ji}|$.  For case (c), we will only consider the flavor-changing couplings of the Majoron and neglect the flavor-conserving ones.  Such processes constrain the products $|c_{V/A}^{ji}||c^{k \ell }_{V/A}|$.  The upper bounds from cases (b) and (c) are given respectively in blocks II and III of Table~\ref{tab:constraints}.
From the above, we see that the muonium-anti-muonium oscillation constrains $|c_{V/A}|$ to be less than around 0.4 if the Majoron scale $f_J$ is 1~TeV, similar in magnitude to the constraints from the $\ell_i \to \ell_j \ell_k \bar \ell_\ell$ decays.  The most stringent constraint on the couplings comes from $\mu \to e J$ with $|c_{V/A}| \alt 3.6\times 10^{-7}$ for $f_J = 1$~TeV.  If one takes $|c_{V/A}| \simeq 10^{-3}$ instead, the best constraint for $f_J$ is $\agt 3000$~TeV.
With improved sensitivity in branching ratio determination, the bounds can be pushed further.

At one loop level, exchanges of Majoron can contribute to $g-2$ of charged leptons and $\ell_i \to \ell_j \gamma$ decays.  The Majoron contribution to $(g-2)_{e,\mu}$ is generally small, giving relatively weak bounds on the couplings, as given in block IV of Table~\ref{tab:constraints}.  Note that because of the opposite deviations, $(g-2)_e$ ($(g-2)_\mu$) constrains the axial (vector) couplings.  Among the $\ell_i \to \ell_j \gamma$ constraints, given in block V, the strongest comes from the $\mu \to e \gamma$ decay: $\sqrt{|c^{e \tau}_{V/A}||c^{\tau \mu }_{V/A}|} < 0.011 ~f_J/\mbox{TeV}$, assuming that flavor-conserving couplings are negligible.  We note in passing that for these loop processes, we have explicitly checked that the same results are obtained by using on-shell current interactions, Eq.~\eqref{eq:on-shell}, and the derivative Majoron couplings, Eq.~\eqref{eq:derivative}.

We now work out the constraint on the coupling between Higgs and Majoron, $\lambda_{hs}$, using the invisible Higgs decay branching ratio bound, ${\rm Br}(h \to {\rm invisible}) <19\%$, from the LHC~\cite{Sirunyan:2018owy}.  This is because $\lambda_{hs}$ can mediate the $h \to JJ$ decay with $\Gamma({h \to JJ })=\lambda_{hs}^2 v^2 \cos^2\theta/32\pi m_{h}$.  Hence, this process contributes to the invisible width of Higgs.   Due to Higgs mixing, the width of the usual SM decay modes will be modified to $\Gamma_{\rm SM} \cos^2\theta$. Using $\Gamma_{\rm SM}=4.07~\mbox{MeV}$~\cite{Heinemeyer:2013tqa} and the modified invisible branching ratio, we obtain a strong constraint of $\lambda_{hs} < 0.014$.

\section{Polarization Asymmetry}
\label{sec:Polarization Asymmetry}

To determine the chiral nature of the Majoron interactions with charged leptons, we propose a novel measurement using the polarizations of the final-state leptons in the $\ell_i \to \ell_j J$ decays.  The polarization 4-vector spinor is $s_i = (\vec n_i\cdot \vec p_i/m_i ,~ \vec n_i + \vec p_i (\vec n_i \cdot \vec p_i)/[m_i(m_i +E_i)])$, where $\vec n_i$ is the polarization of lepton in its rest frame.
For high energy leptons, {\it i.e.}, $E_{i,j} \gg m_{i,j}$, an initially left-handed or right-handed lepton $\ell_{iL, iR}$ can lead to a daughter lepton that is left-handed $\ell_{jL}$ with $\vec n \cdot \vec p = - p$ or right-handed $\ell_{j R}$ with $\vec n \cdot \vec p = p$.  Therefore, there are all four combinations of LL, RR, LR and RL for initial and final lepton polarizations.  The helicity-conserving and -flipping decay rates are given respectively by
\begin{align}
\label{helicity-conserving}
\begin{split}
&
\Gamma_{LL, RR} (\ell_i \to \ell_j J)
\\
&= {m_i m_j\over 64 \pi E_i} \Big[
\Big( |g_{e1}^{ji}|^2 - |g_{e2}^{ji}|^2 \Big) \Big( 2 - {2\over x^2_{ij}} \Big)
\\
& \quad + \Big( |g_{e1}^{ji}|^2 + |g_{e2}^{ji}|^2 \Big)
\Big( {2\over x_{ij}} \ln x_{ij} + {x_{ij}\over 2 } - {1\over 2 x^3_{ij} } \Big)
\\
& \quad \pm {\rm Re} \left(g^{ji}_{e1}g^{ji*}_{e2} \right)
\Big(x_{ij}  -{1\over x_{ij}^3} - {4\over x_{ij}}\ln x_{ij} \Big) \Big]
~,
\end{split}
\end{align}
and
\begin{align}
\label{helicity-flipping}
\begin{split}
&\Gamma_{LR,RL} (\ell_i \to \ell_j J)
\\
&={m_i m_j\over 64 \pi E_i}
\Big[ |g_{e1}^{ji}|^2 + |g_{e2}^{ji}|^2 \mp 2{\rm Re}\left(g^{ji}_{e1}g^{ji*}_{e2} \right) \Big]
\\
&\quad \times \Big(
{ x_{ij} \over 2}  - {1\over 2 x_{ij}^3}
- {2\over x_{ij} }\ln x_{ij} \Big)
~,
\end{split}
\end{align}
where $x_{ij} \equiv m_i/m_j > 1$.   Our result is more general than that given in Ref.~\cite{Kim:1990km}, in which terms proportional to ${\rm Re}(g^{ji}_{e1}g^{ji*}_{e2})$ vanish under their coupling assumption.
More detailed information about deriving the above results is given in the Appendix.

In practice, the polarization of initial-state lepton, presumably produced through collisions, is not easy to determine.  We thus need to average over them.  But the polarization of final-state lepton can be measured.  Taking $\tau \to \mu J$ as an example, we define the polarization asymmetry
\begin{align}
\begin{split}
A
&\equiv
{\Gamma_{LL} + \Gamma_{RL} - \Gamma_{LR} - \Gamma_{RR} \over  \Gamma_{LL} + \Gamma_{RL} + \Gamma_{LR} + \Gamma_{RR} }
\approx
-2 {{\rm Re}(c_V^{\mu\tau}c_A^{\mu\tau *}) \over |c_V^{\mu\tau}|^2 + |c_A^{\mu\tau}|^2}
~,
\end{split}
\end{align}
where the second expression neglects terms of order $m_\mu/m_\tau$.  This quantity probes the Majoron interaction in more detail.

As an explicit example, consider the $\tau^+ \to \mu^+ J$ decay~\footnote{In order to determine the muon polarization, as required in our polarization asymmetry, it is preferred to study $\mu^+$ because $\mu^-$ may have reactions with surrounding matter via $\mu^- p \to \nu_\mu n$.} and we measure the longitudinal polarization of $\mu^+$, denoted by $P_L$.  Since $A =2P_L -1$, the precision on the asymmetry measurement depends on how accurately $P_L$ can be determined.  Assuming $c_V = - c_A$, the final-state anti-muon is dominantly right-handed and $A \approx 1$, regardless of how $\tau$ is polarized.  This value would be reduced by about 3\% if corrections from $(m^2_\mu/m_\tau^2)\ln(m_\mu/m_\tau)$ and $m^2_\mu/m^2_\tau$ are taken into account.  Currently, ${\rm BR}(\tau \to \mu J)$ is constrained to be less than $5.7\times 10^{-3}$.  To have an estimate about the precision one can reach for $A$, let's take the branching ratio to be $10^{-4}$, which is well below the current bound, as an example.  Given the fact that BELLE-II will produce in total about 45 billion $\tau^+\tau^-$ pairs~\cite{belle2-tau}, one expects to observe ${\cal O}(10^6)$ $\tau^+ \to \mu^+ J$ decays.  The polarization of $\mu$ can be obtained from the $\mu^+ \to e^+ \nu_e \overline\nu_\mu$ decay by measuring the energy spectrum of positrons.  Without considering the detection efficiency, the statistical error on the muon polarization determination is seen to be at the per mille level.  If the decay branching ratio is different, the statistical error is then scaled by a factor of $\sqrt{10^{-4}/{\rm Br}(\tau\to \mu J)}$.

Another significant background source is the Michel decay $\tau^+ \to \mu^+ {\nu}_\mu \overline{\nu}_\tau$ with a branching ratio of $17.39\%$.  As this is a three-body decay while our signal process is a two-body decay, one can impose a cut on the kinematic variable $x \equiv 2E_\mu / m_\tau$ in the rest frame of $\tau$ to remove most of the background.  Assuming that experiment can impose the cut $0.99 \le x \le 1$, then the branching ratio of the background is reduced to $\simeq 3.5 \times 10^{-3}$.  Assuming again that $\operatorname{Br}(\tau \rightarrow \mu J) = 10^{-4}$, we will expect a statistical error of about $0.3\%$ from BELLE-II data.  Finally, the systematic error on the muon $P_L$ measurement at Spin Muon Collaboration had been estimated to be $\sim 3\%$~\cite{smc}, making the total error at a few percent level.  We therefore encourage our experimental colleagues to carry out such an analysis.

We note in passing that, in fact, the polarization asymmetry can also be obtained from $\tau$ decays at rest, in which case $A = - 2 {\rm Re}(c_V c_A^*)/(|c_V|^2+|c_A|^2)$.  Therefore, one can determine the chirality of Majoron interaction from $\tau$ decays at low speeds, such as those produced at threshold by BES-III, where 600 million $\tau$ pairs have been obtained~\cite{bes3-tau}, and the future Super Tau-Charm Factory, where a few billion $\tau$ pairs per year are expected.

It may be tempting to use $\mu \to e J$ to determine the corresponding $A$ by measuring muon or electron polarization since a high-luminosity muon beam will be available in $\mu$-$e$ conversion experiment at COMET and Mu2e.  This turns out to be rather difficult for several reasons.  The branching ratio of $\mu \to e J$ is bounded to be smaller than that of $\tau \to \mu J$ (by a factor of $\sim 10^{-3}$) and, hence, can offset the gain from high-luminosity muon beam for the $\mu$-$e$ conversion experiment.  Secondly, the energy of the electron will be half of the $\mu$-$e$ conversion experiment, and only a very small fraction of $\mu \to e J$ decays resides in the signal region on target.  It is therefore very difficult to measure such a process at COMET and Mu2e~\cite{Calibbi:2020jvd}.  If the polarization information of the electron is further required, it would pose more difficulty as it does not decay.  Alternatively, one may consider using polarized initial-state muons to construct an analogous asymmetry by summing over the final-state electron helicities.  However, the corresponding asymmetry is identically zero.

\section{Conclusion}
\label{sec:Conclusion}

We have used the type-III seesaw model as an explicit example to motivate a Majoron with flavor-changing interactions with SM fermions, though our study is largely model-independent.  We have examined existing major experimental constraints on the Majoron, including the ratios of the couplings, $c_{V/A}^{\ell'\ell}$, the Majoron decay constant, $f_J$, and the coupling between the Higgs boson and the Majoron, $\lambda_{hs}$.  Finally, we propose an experimental observable, the polarization asymmetry, in the $\ell_i \to \ell_j J$ decays.  Using the $\tau \to \mu J$ decay as an example, we conclude that through the measurement of $A$, it is a promising channel to probe the chiral nature of Majoron couplings with the charged leptons.
As a final remark, if the Majoron is replaced by a Majoron-like particle with a finite mass, one can carry out a similar analysis so long as the decays are kinematically allowed.

\acknowledgements

XGH thanks Haibo Li for many interesting discussions.  This work was supported in part by NSFC (Grants 11735010, 11975149, 12090064), by Key Laboratory for Particle Physics, Astrophysics and Cosmology, Ministry of Education, and Shanghai Key Laboratory for Particle Physics and Cosmology (Grant No. 15DZ2272100), and in part
by the MOST (Grant Nos.~108-2112-M-002-005-MY3 and 109-2112-M-002-017-MY3

\section{Appendix}

From the interaction Lagrangian given in Eq.~\eqref{eq:on-shell}, we get the decay matrix element
\begin{equation}
	-i{\cal M} = \frac{1}{2} i \overline{\ell}_j (g_{e1}+g_{e2} \gamma_5) \ell_i.
\end{equation}
\begin{widetext}
The absolute-squared matrix element is given by
\begin{align}
\begin{split}
 |{\cal M}|^2 =& \frac{1}{16} \mbox{Tr} \left[ (\slashed{p}_j + m_j ) (1 + \gamma_5 \slashed{s}_j) (g_{e1}+g_{e2} \gamma_5) 
(\slashed{p}_i + m_i) (1 + \gamma_5 \slashed{s}_i)
(g^*_{e1}-g^*_{e2} \gamma_5) \right]
\\
=&
\frac{1}{4}|g_{e1}|^2 \left[ (p_j \cdot p_i + m_j m_i ) (1 - s_j \cdot s_i ) + p_j \cdot s_i \;p_i \cdot s_j \right]
\\
&+ \frac{1}{4}|g_{e2}|^2 \left[ (p_j \cdot p_i - m_j m_i ) (1 + s_j \cdot s_i ) - p_j \cdot s_i\; p_i \cdot s_j \right]\\
&+ \frac{1}{2} {\rm Re}(g_{e1} g^*_{e2}) (m_i p_j \cdot s_i - m_j p_i \cdot s_j)
~.
\end{split}
\end{align}
Using the polarization 4-vector spinor $s_i^\mu$ defined in the main text, we get for different helicity combinations that
\begin{align}
\begin{split}
|{\cal M}|_{\ell_i^{L,R} \rightarrow \ell_j^{L,R}}^2
=& \frac{1}{4} |g_{e1}|^2 \left[ \frac{1}{2} (m_i + m_j )^2
\left( 1- \frac{E_i E_j}{2 m_i m_j |\mathbf{p}_{i}| |\mathbf{p}_{j}|} [m^2_i + m^2_j - 2 m_i m_j (A - \frac{m_i m_j}{E_i E_j} ) ] \right) \right.
\\
&\qquad
\left. + \frac{E_i E_j}{4 m_i m_j |\mathbf{p}_{i}| |\mathbf{p}_{j}|}[4 m^2_i m^2_j+ (m^2_i + m^2_j)^2 - 2 m_i m_j (m^2_i + m^2_j)A] \right]
\\
&+\frac{1}{4}|g_{e2}|^2 \left[ \frac{1}{2} (m_i - m_j )^2
\left( 1 + \frac{E_i E_j}{2 m_i m_j |\mathbf{p}_{i}| |\mathbf{p}_{j}|} [m^2_i + m^2_j - 2 m_i m_j (A - \frac{m_i m_j}{E_i E_j} ) ] \right) \right.
\\
&\qquad
\left. - \frac{E_i E_j}{4 m_i m_j |\mathbf{p}_{i}| |\mathbf{p}_{j}|}[4 m^2_i m^2_j+ (m^2_i + m^2_j)^2 - 2 m_i m_j (m^2_i + m^2_j)A] \right] \\
& \pm \frac{1}{2} {\rm Re} (g_{e1} g^*_{e2}) \left(\frac{1}{2 |\mathbf{p}_{i}|} [2 E_j m^2_i-(m^2_i+m^2_j)E_i]
- \frac{1}{2 |\mathbf{p}_{j}|} [2 E_i m^2_j-(m^2_i+m^2_j)E_j] \right)
\end{split}
\end{align}
and
\begin{align}
\begin{split}
|{\cal M}|_{\ell_i^{R,L} \rightarrow \ell_j^{L,R}}^2
=& \frac{1}{4}|g_{e1}|^2 \left[ \frac{1}{2} (m_i + m_j )^2
\left( 1+ \frac{E_i E_j}{2 m_i m_j |\mathbf{p}_{i}| |\mathbf{p}_{j}|} [m^2_i + m^2_j - 2 m_i m_j (A - \frac{m_i m_j}{E_i E_j} ) ] \right) \right.
\\
&\qquad
\left. - \frac{E_i E_j}{4 m_i m_j |\mathbf{p}_{i}| |\mathbf{p}_{j}|}[4 m^2_i m^2_j+ (m^2_i + m^2_j)^2 - 2 m_i m_j (m^2_i + m^2_j)A] \right]
\\
&+\frac{1}{4}|g_{e2}|^2 \left[ \frac{1}{2} (m_i - m_j )^2
\left( 1 - \frac{E_i E_j}{2 m_i m_j |\mathbf{p}_{i}| |\mathbf{p}_{j}|} [m^2_i + m^2_j - 2 m_i m_j (A - \frac{m_i m_j}{E_i E_j} ) ] \right) \right.
\\
&\qquad
\left. + \frac{E_i E_j}{4 m_i m_j |\mathbf{p}_{i}| |\mathbf{p}_{j}|}[4 m^2_i m^2_j+ (m^2_i + m^2_j)^2 - 2 m_i m_j (m^2_i + m^2_j)A] \right] \\
& \pm \frac{1}{2} {\rm Re} (g_{e1} g^*_{e2}) \left(-\frac{1}{2 |\mathbf{p}_{i}|} [2 E_j m^2_i-(m^2_i+m^2_j)E_i]
- \frac{1}{2 |\mathbf{p}_{j}|} [2 E_i m^2_j-(m^2_i+m^2_j)E_j] \right)
~,
\end{split}
\end{align}
where
\begin{equation}
A \equiv\left(\frac{m_{i} E_{j}}{m_{j} E_{i}}+\frac{m_{j} E_{i}}{m_{i} E_{j}}\right)
\end{equation}
\end{widetext}
and $E_{i,j}$ and ${\bf p}_{i,j}$ are the energy and 3-momentum associated with $\ell_i$ and $\ell_j$, respectively.
The differential decay rate of the lepton flavor-changing process is given by
\begin{equation}
\frac{d \Gamma}{d E_{j}}=\frac{1}{16 \pi E_{i}\left|\mathbf{p}_{i}\right|}|\mathcal{M}|^{2}
~.
\end{equation}
To get the total decay rate of the $\ell_i \to \ell_j J$ process, we must integrate over the allowed energy range for the final state lepton $\ell_j$,
\begin{equation}
E_{j}^{\rm max,min} =
\frac{E_{i}}{2}\left(1+\frac{1}{x_{i j}^{2}}\right) \pm \frac{\left|\mathbf{p}_{i}\right|}{2}\left(1-\frac{1}{x_{i j}^{2}}\right)
~,
\end{equation}
where $x_{i j} \equiv m_{i} / m_{j}$ and in the lab frame with $E_{i} \gg m_{i}$, the energy range becomes $E_{i} / x_{i j}^{2} \leq E_{j} \leq E_{i}$.
After integrating over the energy range, we obtain the decay rates for the helicity-conserving and -flipping processes given in Eqs.~\eqref{helicity-conserving} and \eqref{helicity-flipping}, respectively.


\end{document}